\documentclass[]{spie}  

\usepackage{amsmath,amsfonts,amssymb}
\usepackage{graphicx}
\usepackage[colorlinks=true, allcolors=blue]{hyperref}

\title{METIS high-contrast imaging: from final design to manufacturing and testing}

\author[a]{Olivier Absil}
\author[b]{Matthew Kenworthy}
\author[a]{Christian Delacroix}
\author[a]{Gilles Orban de Xivry}
\author[c,a]{Lorenzo~König}
\author[d,a]{Prashant Pathak}
\author[e,b]{David Doelman}
\author[f,b]{Emiel Por}
\author[b]{Frans Snik}
\author[g]{Joost~van~den~Born}
\author[h]{Faustine Cantalloube}
\author[h]{Alexis Carlotti}
\author[i,n]{Benjamin Courtney-Barrer}
\author[j]{Pontus Forsberg}
\author[j]{Mikael Karlsson}
\author[k]{Thomas Bertram}
\author[k]{Roy van Boekel}
\author[g]{Dennis Dolkens}
\author[k]{Markus Feldt}
\author[l]{Adrian M.\ Glauser}
\author[m]{Eric Pantin}
\author[l]{Sascha~P.~Quanz}
\author[b]{Felix Bettonvil}
\author[b]{Bernhard Brandl}
\affil[a]{STAR Institute, Universit\'e de Li\`ege, All\'ee du Six Ao\^{u}t 19c, B-4000 Li\`ege, Belgium}
\affil[b]{Leiden Observatory, Leiden University, P.O. Box 9513, 2300 RA Leiden, The Netherlands}
\affil[c]{Jet Propulsion Laboratory, California Institute of Technology, 4800 Oak Grove Dr., Pasadena, CA 91109, USA}
\affil[d]{Department of SPASE, Indian Institutes of Technology Kanpur, 208016, UP, India}
\affil[e]{SRON Netherlands Institute for Space Research, Niels Bohrweg 4, 2333 CA, Leiden, The Netherlands}
\affil[f]{Space Telescope Science Institute (STScI), 3700 San Martin Dr, Baltimore MD, 21218, USA}
\affil[g]{NOVA Optical and Infrared Instrumentation Group, P.O. Box 2, 7990 AA Dwingeloo, The Netherlands}
\affil[h]{Universit\'e Grenoble Alpes, CNRS, IPAG, 38000 Grenoble, France}
\affil[i]{European Southern Observatory, Alonso de C\'ordova 3107, Vitacura, Santiago, Chile}
\affil[j]{Department of Materials Science and Engineering, Uppsala University, L\"agerhyddsv\"agen 1, 751\,03 Uppsala, Sweden}
\affil[k]{Max-Planck-Institut f\"{u}r Astronomie, K\"{o}nigstuhl 17, 69117 Heidelberg, Germany}
\affil[l]{ETH Z\"urich, Institute for Particle Physics \& Astrophysics, Wolfgang-Pauli-Str.~27, 8093 Z\"urich, Switzerland}
\affil[m]{Universit\'e Paris-Saclay, Universit\'e Paris Cit\'e, CEA, CNRS, AIM, 91191 Gif-sur-Yvette, France}
\affil[n]{Research School of Astronomy and Astrophysics, Australian National University, Canberra 2611, Australia}

\authorinfo{Further author information: (Send correspondence to O.A.)\\O.A.: E-mail: olivier.absil@uliege.be}

\begin{document} 
\maketitle

\begin{abstract}
The Mid-infrared ELT Imager and Spectrograph (METIS) is one of the first-generation scientific instruments for the ELT, built under the supervision of ESO by a consortium of research institutes across and beyond Europe. Designed to cover the 3 to 13~$\mu$m wavelength range, METIS had its final design reviewed in Fall 2022, and has then entered in earnest its manufacture, assembly, integration, and test (MAIT) phase. Here, we present the final design of the METIS high-contrast imaging (HCI) modes. We detail the implementation of the two main coronagraphic solutions selected for METIS, namely the vortex coronagraph and the apodizing phase plate, including their combination with the high-resolution integral field spectrograph of METIS, and briefly describe their respective backup plans (Lyot coronagraph and shaped pupil plate). We then describe the status of the MAIT phase for HCI modes, including a review of the final design of individual components such as the vortex phase masks, the grayscale ring apodizer, and the apodizing phase plates, as well as a description of their on-going performance tests and of our plans for system-level integration and tests. Using end-to-end simulations, we predict the performance that will be reached on sky by the METIS HCI modes in presence of various environmental and instrumental disturbances, including non-common path aberrations and water vapor seeing, and discuss our strategy to mitigate these various effects. We finally illustrate with mock observations and data processing that METIS should be capable of directly imaging temperate rocky planets around the nearest stars.  
\end{abstract}


\section{INTRODUCTION}
\label{sec:intro}  

\begin{figure}[t]
\begin{center}
\begin{tabular}{c} 
\includegraphics[height=10cm]{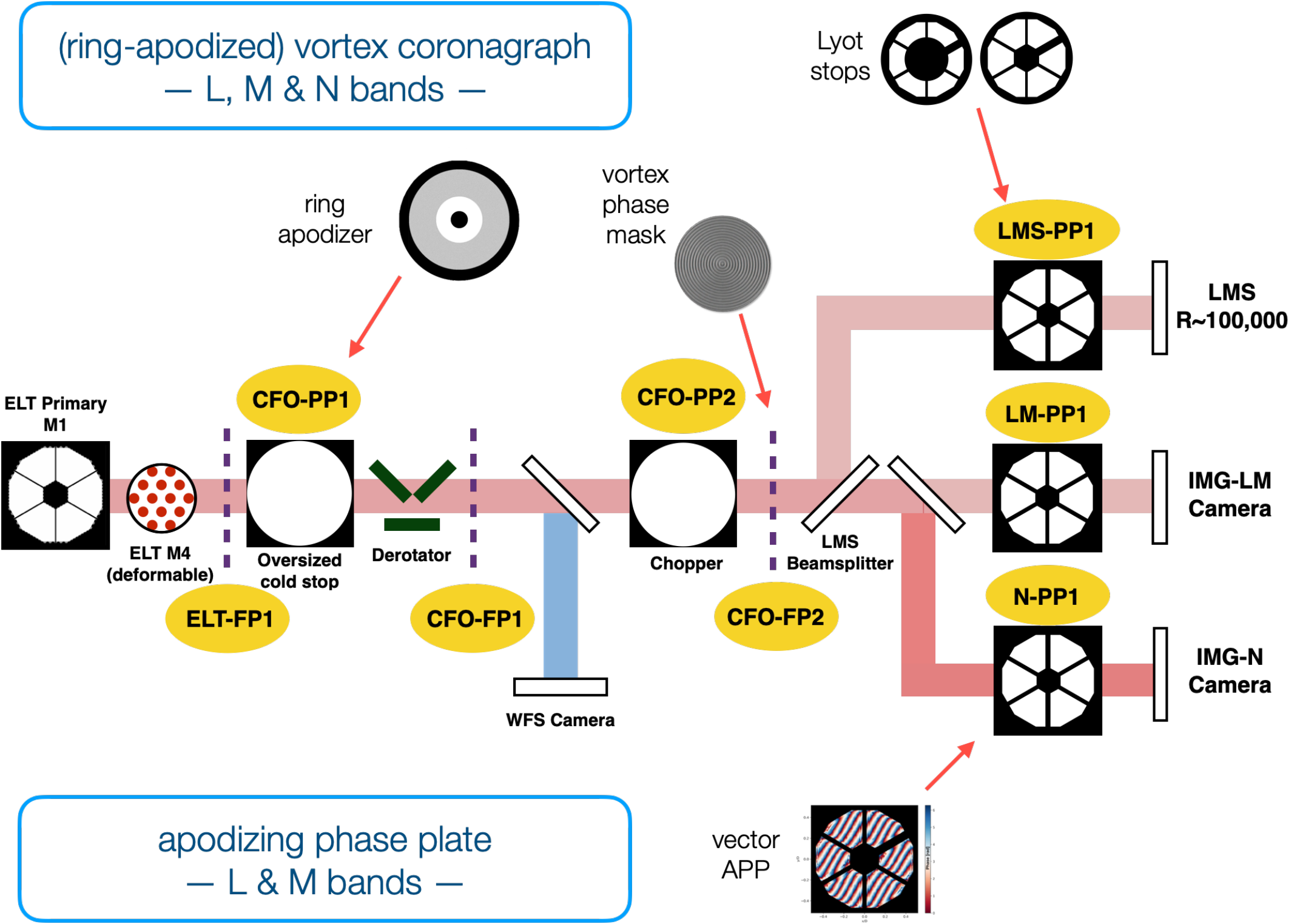}
\end{tabular}
\end{center}
\caption[tube]{\label{fig:tube} Schematic representation the METIS design, and of the placement of HCI components within METIS for the two baseline coronagraphic concepts (vortex coronagraph and apodized phase plate). Pupil and focal planes are highlighted with orange ellipses, and some of the main functionalities of the instrument are displayed.}
\end{figure}

The mid-infrared ELT imager and spectrograph (METIS)\cite{Brandl21} is one of the first-generation instruments of the Extremely Large Telescope. METIS recently passed its final design review, and is expected to see first light in 2029\cite{Brandl24}. Specifically designed to deliver high-contrast imaging (HCI) capabilities, METIS features a high-performance adaptive optics module and advanced coronagraphic concepts that can be combined with imaging cameras covering the 3--13~$\mu$m range, as well as with a high-resolution (R=100,000) integral field spectrograph operating from 3 to 5~$\mu$m. The METIS instrument is divided into a series of sub-systems, all located inside the METIS cryostat:
\begin{itemize}
    \item The CFO (common fore-optics) implements a series of important functions such as derotation and chopping. It gives access to rotating wheel mechanisms in a pupil plane (CFO-PP1) upstream of the derotator and in a focal plane (CFO-FP2) downstream of the chopper, where coronagraphic elements can be implemented. Another focal plane (CFO-FP1) and pupil plane (CFO-PP2) are respectively used up by a pupil stabilization mirror and by the chopper.
    \item The SCAO (single-conjugate adaptive optics) implements an infrared pyramid wavefront sensor operating in the H and K bands, which is used to control the ELT-M4 and M5 adaptive mirrors\cite{Bertram24}. SCAO is also used to control the pupil stabilization mirror located in the first focal plane of CFO. The beam feeding SCAO is picked up just after this focal plane, upstream of the chopper.
    \item The IMG (imager), located after the second focal plane of the CFO, includes two infrared cameras covering respectively the LM bands (IMG-LM) and the N band (IMG-N). Both arms include a dedicated pupil plane (IMG-LM-PP1 and IMG-N-PP1), where cold stops and/or coronagraphic elements can be placed.
    \item The LMS (LM spectrograph) is fed by sliding a mirror or beamsplitter just after the second focal plane of the CFO. If the beamsplitter is used, 10\% of the light will simultaneously be fed to the IMG. After passing through a pupil plane (LMS-PP1), where coronagraphic elements can be placed, the beam enters a field stop before being presented to the image slicer and propagating through the rest of the spectrograph.
\end{itemize}
METIS also includes a warm calibration unit (WCU) outside of its cryostat. A schematic representation of the various pupil planes and image planes available in METIS is given in Fig.~\ref{fig:tube}.

METIS will address various science cases, ranging from our solar system to distant galaxies, but is more specifically optimized to study nearby planetary systems and planetary formation using a combination of HCI and high-resolution spectroscopy (HRS).

\section{THE METIS HIGH-CONTRAST IMAGING MODES}
\label{sec:metis}

METIS is expected to bring major breakthroughs in exoplanet science thanks to the exquisite angular resolution and sensitivity enabled by the 39-m ELT aperture. To take full advantage of this situation, a vortex coronagraph\cite{Mawet05} is implemented in METIS to reach the smallest inner working angle (IWA) while maintaining a high throughput. A single-grating vector apodizing phase plate (vAPP)\cite{Doelman21}, creating two conjugated PSFs with dark holes on either side of the star, is also implemented for more robustness to aberrations, at the expense of a larger IWA and lower throughput.

    \subsection{Vortex coronagraphy}

Our implementation of vortex coronagraphy is based on the annular groove phase mask (AGPM)\cite{Mawet05} concept, which uses the form birefringence of a concentric subwavelength grating to create a spatially variant half-wave plate defining a helical phase ramp through geometric (Pancharatnam-Berry) phase. The implementation of this specific vortex phase mask (VPM) using reactive ion etching on a diamond substrate has already demonstrated its performance in the thermal infrared both in the lab\cite{Delacroix13,Forsberg24} and on sky\cite{Absil16}. The METIS CFO-FP2 focal-plane wheel will host three different types of VPMs, covering respectively the L band (VPM-L), the M band (VPM-M), and the N band (VPM-N), with typical operating bandwidths ranging from about 30\% (L and M bands) to 50\% (N band). The VPM gratings have been specifically designed to cover these wide bands with high coronagraphic performance, following the same methodology as in previous works\cite{VargasCatalan16}. In order not to drive the overall HCI performance, the VPMs need to achieve rejection rates higher than 500:1 on average over the L-M bands, and higher than 100:1 on average over the N band. We also require the rejection rates to be better than 100:1 at any given wavelength in L-M bands, and better than 20:1 at any given wavelength in N band. Such performance is compatible with the proposed operating bandwidths. In the case of N-band observations, three identical VPM-N will be placed side-by-side in the CFO-FP2 focal-plane wheel to enable three-point chopping patterns for efficient background subtraction.

\begin{figure}[t]
\begin{center}
\begin{tabular}{c} 
\includegraphics[width=\textwidth]{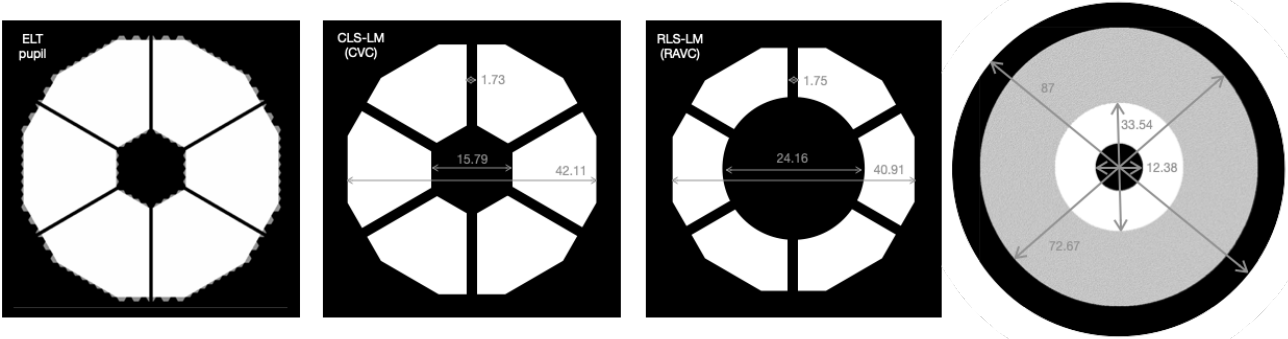}
\end{tabular}
\end{center}
\caption[lyot]{\label{fig:lyot} \textit{Left.} ELT input pupil, illustrating our dodecagonal/hexagonal all-glass approximation (white) compared to the full input pupil (gray). \textit{Middle.} Design of the Lyot stops associated with the classical (middle left) and ring-apodized (middle right) vortex modes, as implemented at IMG-LM-PP1. Sizes are in given in mm, with the ELT nominal diameter corresponding to 45~mm in this pupil plane. \textit{Right.} Design of the ring apodizer (RAP). Sizes are in given in mm, with the ELT nominal diameter corresponding to 68~mm in this pupil plane. The relative dimensions of all masks with respect to the input ELT pupil are respected to illustrate the amount of oversizing or undersizing in the various edges.}
\end{figure}

In order to perform coronagraphic observations, the VPMs need to be associated with Lyot stops in a downstream pupil plane, resulting in what we call the classical vortex coronagraph (CVC) observing mode in METIS. The Lyot stops are implemented inside the scientific cameras, respectively in the IMG-LM-PP1, IMG-N-PP1, and LMS-PP1 pupil plane wheels. They are designed to account for the effects of pupil instabilities and pupil blurring, which require their outer edge to be undersized, while their central obscuration and spiders need to be oversized, compared to the input ELT pupil. The amount of undersizing/oversizing is driven both by the stellar extinction performance and by need to properly reject the background emission within the considered field of view (the Lyot stops also act as cold stops). The undersizing of the outer pupil edge is computed relative to the largest inscribed regular dodecagon within the ELT-M1 pupil (38177~mm in diagonal in ELT-M1 space), while the oversizing of the central obscuration is computed relative to the smallest circumscribed hexagon around the ELT-M1 central obscuration (11608~mm in diagonal in ELT-M1 space). The size difference of the Lyot stops relative to the ELT pupil typically corresponds to about 2\% to 3\% of the ELT nominal pupil diameter (38542~mm in ELT-M1 space). In addition, because a significant amount of residual stellar light is located around the central obscuration in a pupil plane downstream the VPM, the size of the central obscuration has been increased by 9\% of the ELT nominal diameter -- a value that results from a trade-off between throughput and stellar light cancellation efficiency. The design of the ``classical'' Lyot stop (referred to as CLS) for the CVC mode is represented in Figure~\ref{fig:lyot} in the case of IMG-LM-PP1, where the ELT nominal pupil diameter corresponds to about 45~mm. In practice, five of the six spider arms will be made slightly thinner thanks to the latest evolution in the design of the ELT spider arms. Also, as explained in Section~\ref{sub:ncpa}, the thickness of one or two of these spider arms (or part thereof) will be made thicker to enable focal-plane wavefront sensing using an asymmetric Lyot sensor concept. The overall throughput of the CVC mode relative to the input ELT pupil at 4~$\mu$m, taking into account the VPM transmission, the Lyot stop transmission, and the loss in encircled energy in a 1-FWHM aperture due to the Lyot stop shape, amounts to about 53\% at 3.8~$\mu$m, not taking into account the off-axis transmission profile of the CVC mode. It would reduce to about 44\% for an actual off-axis companion at $5\lambda/D$ from the optical axis. This is to be compared with a typical throughput of 70\% for the standard imaging mode (this value accounts for the throughput of the cold stop and associated loss in encircled energy).

In addition to the CVC mode, METIS includes a more aggressive coronagraphic mode also using the VPM as a phase mask, which aims to further reduce the amount of stellar leakage compared to the CVC mode at the expense of throughput. This mode is based on the concept of the ring-apodized vortex coronagraph (RAVC)\cite{Mawet13}, which compensates for the harmful diffractive effect of the central obscuration on the vortex coronagraph rejection rate by introducing a grayscale apodizer (referred to as ring apodizer, or RAP) in an upstream pupil plane. When properly designed in terms of throughput and geometry, this compensation would be perfect for an annular pupil in the absence of optical aberrations. In our case, the performance of the RAVC is limited by the dodecagonal and hexagonal shapes of the outer and inner edges of the ELT-M1 pupil, as well as by the presence of the spiders. Because it is placed in a pupil plane (CFO-PP1) upstream of the derotator, where the pupil is not stabilized in terms of clocking, the RAP cannot be designed to (even partly) compensate for these non-radial features of the ELT input pupil. Nevertheless, the theoretical rejection rate provided by the RAVC can be improved by a factor around 10 compared to the CVC mode, which will be particularly useful for the brightest targets where speckle noise is driving the performance budget. The higher rejection rate and lower throughput of the RAVC have been specifically designed to allow the observations of $\alpha$~Cen~A (magnitude $L=-1.6$) without the need of a neutral density. The design of the RAP is illustrated in Figure~\ref{fig:lyot}. The grayscale area, producing a 62.5\% transmission, is oversized with respect to the input ELT pupil, to make sure that the ELT pupil image always remains fully inscribed inside the outer edge of the RAP even in presence of pupil drifts. The overall throughput of the RAVC in terms of encircled energy at 3.8~$\mu$m for an off-axis companion at $5\lambda/D$ is about 14\% relative to the input ELT pupil.

    \subsection{Apodizing phase plate coronagraphy}

While the CVC is supposed to be the workhorse HCI mode in METIS, the achievable performance of the vortex coronagraph strongly depends on the quality and stability of the wavefront that will be delivered by the ELT combined with the METIS adaptive optics system. In particular, a charge-2 VPM (like the ones used in METIS) is known to be particularly sensitive to pointing errors, including the high-frequency contribution of wind shake on the ELT structure. While these effects have been taken into account in the end-to-end simulations used to assess the performance of the various HCI modes (see Sect.~\ref{sec:e2e}), the early performance of the ELT and of the METIS adaptive optics system remains a significant risk for the early operations of the METIS HCI modes. For this reason, a more robust HCI concept based on a single pupil plane element has been added to the METIS observing mode. The performance of a pupil-plane coronagraph is (to the first order) insensitive to tip-tilt errors caused by vibrations or residual wavefront errors from the adaptive optics system. Moreover, tip-tilt insensitivity is very relevant for thermal infrared observations, where chopping/nodding are required to remove the background. Considering its technological maturity and high performance, the vector apodiziding phase plate (vAPP)\cite{Doelman21} was selected as a baseline for pupil-plane coronagraphy in METIS.

\begin{figure}[t]
\centering 
\includegraphics[width=\textwidth]{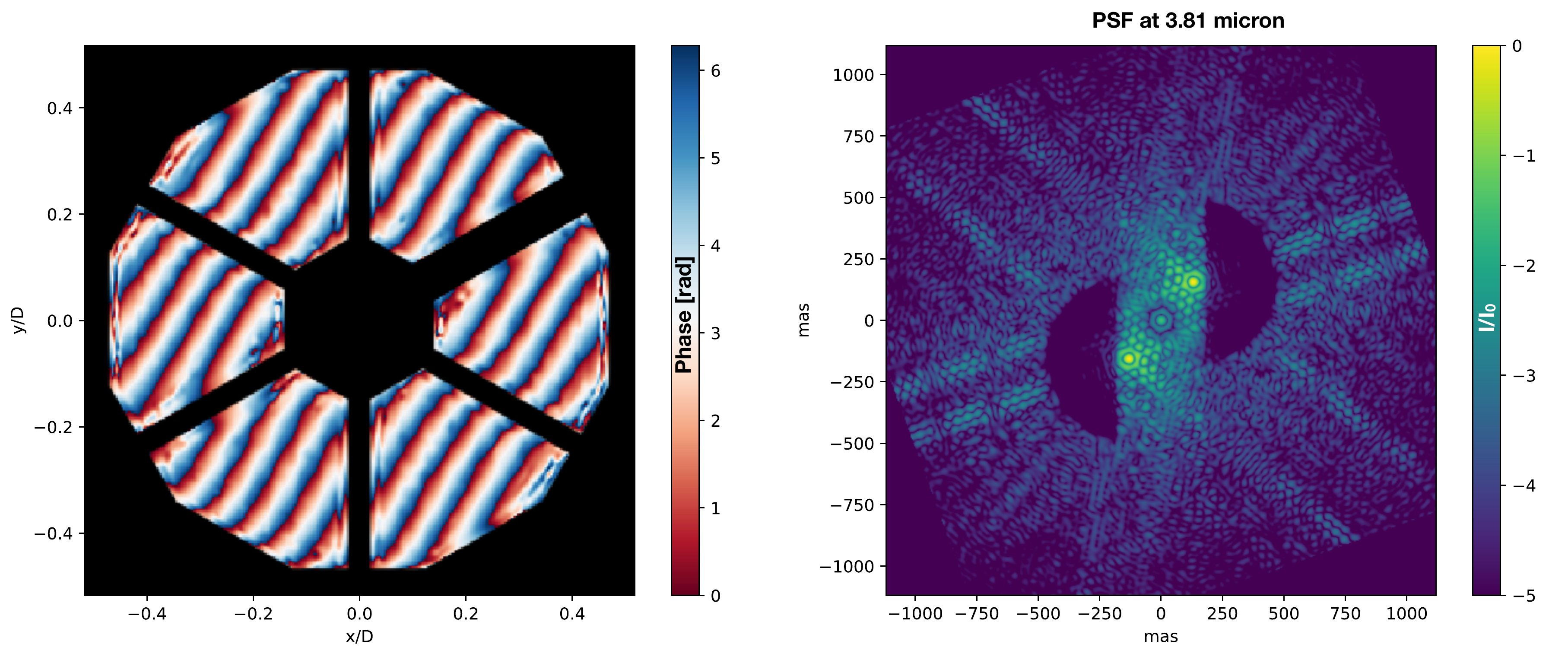}
\caption{APP-IMG phase pattern and resultant monochromatic PSF. \textit{Left.} Phase of the vAPP as a function of position in the METIS cold stop pupil. \textit{Right.} The two (monochromatic) coronagraphic PSFs formed by this phase pattern, located on either side of a central leakage term of about 1\% intensity relative to the input on-axis starlight.}
\label{fig:app-img}
\end{figure}

The vAPP is a transparent pupil plane element that modifies the phase in the pupil plane to create regions in the PSF where the star light is suppressed, so-called dark zones. The vAPP uses the geometric (Pancharatnam-Berry) phase to generate two coronagraphic PSFs that have dark zones on opposite sides of the point spread function, for opposite circular polarization states. Adding a polarization grating to this device enables to spatially separate the two PSFs in the focal plane, as illustrated in Figure~\ref{fig:app-img}. Because their technology has never been tested at wavelengths beyond 5~$\mu$m, and because their lower throughput (compared to the CVC) is not very appropriate for N-band observations, vAPPs are only implemented in the IMG-LM and the LMS. The vAPPs will be placed in the first pupil plane of these two sub-systems, where a high-quality, derotated and stabilized pupil image is available. As for the Lyot stops, the vAPPs will also serve as cold stops, by integrating an opaque metal mask.

The design of the vAPPs was optimized as a trade-off between various competing performance metrics: depth of the dark zones, Strehl ratio, inner working angle (IWA), and outer working angle (OWA). This trade-off also had to take into account the wavelength smearing due to the wavelength-dependent behaviour of the polarization grating, and the specific geometry of the METIS pupil and of the focal-plane detector. The main design guidelines are summarized below in the case of the vAPP that will be installed in the IMG-LM (referred to as APP-IMG):
\begin{itemize}
    \item the center of the two APP-IMG PSFs shall be offset by $10\lambda/D$ from the on-axis leakage term, with an angle of 50$^{\circ}$ to reduce the effect of wavelength smearing as much as possible, and ensure that both coronagraphic PSFs are located within the well-corrected zone by the ELT-M4 deformable mirror;
    \item the dark holes shall be extended along the Hawaii 2RG detector rows (perpendicular to the read-out channels), and the maximum PSF intensity in the detector rows including the dark holes shall be less than $5\times 10^{-3}$ of the peak in the non-coronagraphic PSF;
    \item the extension of the dark holes along the detector columns (read-out channels) shall be defined by an IWA of $2.5\lambda/D$ and an OWA of $20\lambda/D$;
    \item the dark hole depth shall be $<10^{-3.5}$ at the IWA, and gradually go down to a dark hole depth of $10^{-5}$ beyond $5\lambda/D$ (note: this also includes light from the conjugated PSF, which should not perturb the considered dark hole);
    \item the APP-IMG PSF shall have a Strehl ratio larger than 60\%.
\end{itemize}
The resulting design of the phase pattern is illustrated in Figure~\ref{fig:app-img}, together with the resulting PSF. One spider arm of the vAPP pupil mask was purposely made thicker to enable focal-plane wavefront sensing\cite{Bos19} (see also Sect.~\ref{sub:ncpa}). In the case of the LMS, the dark zone of the vAPP was designed to match the size of the image slicer as much as possible, so that the integral field spectrograph receives as few stellar photons as possible. The overall throughput of the APP modes at 3.8~$\mu$m, taking into account the intrinsic transmission of the vAPP, the reduction in encircled energy, and the splitting of the beam into two parts, amounts to about 14\% relative to the input ELT pupil.

    \subsection{HCI back-up modes}

\begin{figure}[t]
\centering 
\includegraphics[width=4.7cm]{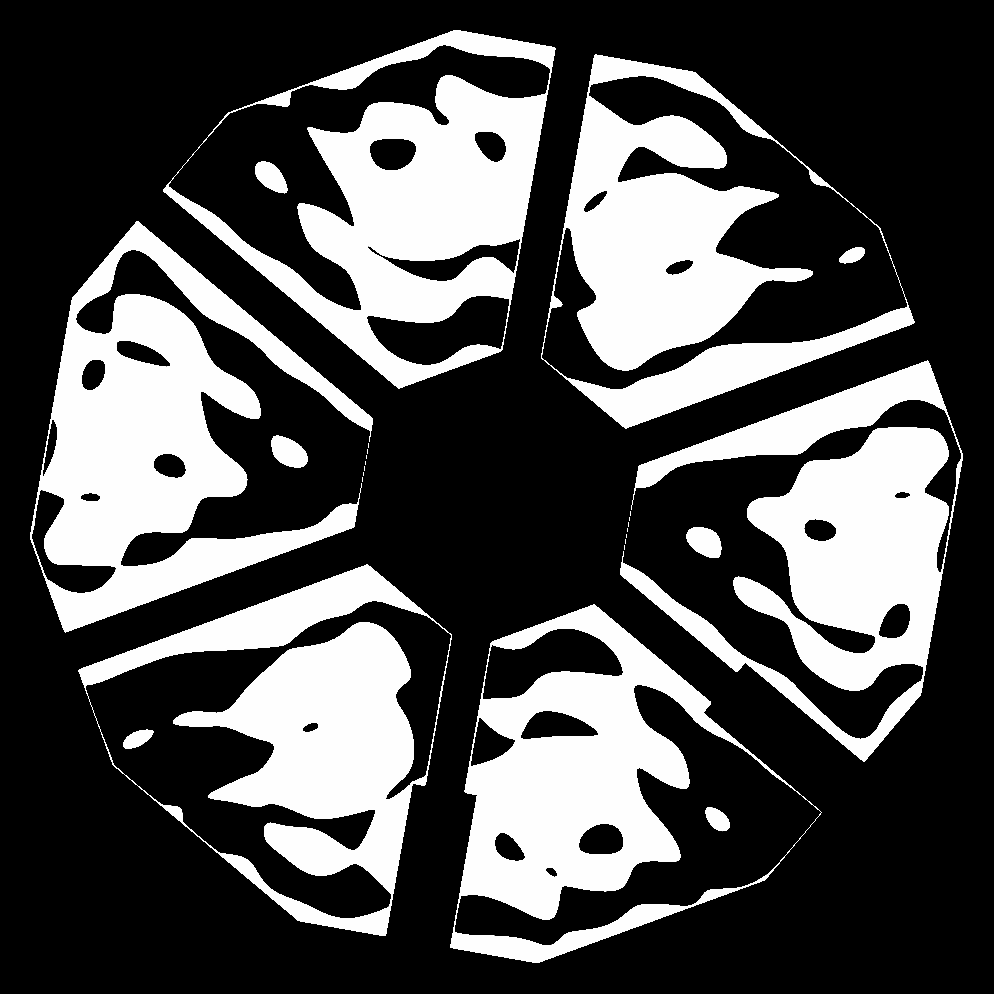}
\includegraphics[width=5.9cm]{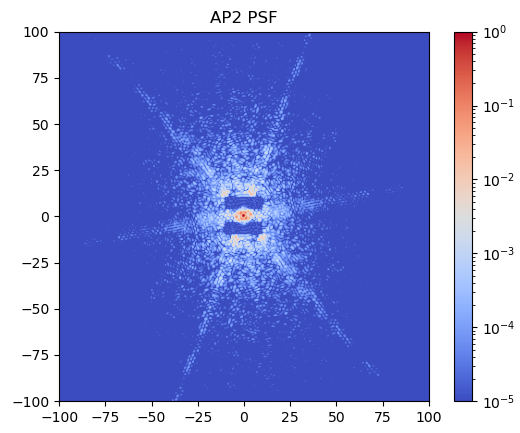}
\includegraphics[width=5.8cm]{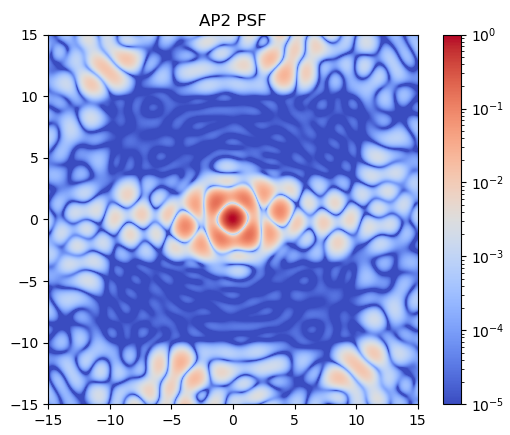}
\caption{\textit{Left.} Design of the shaped pupil plate (SPP), considered as a backup option for the vAPP mode. \textit{Middle \& right.} Illustrations of the associated monochromatic PSF, at two levels of zoom.}
\label{fig:spp}
\end{figure}

Manufacturing high-technology coronagraphic devices such as the VPMs and the vAPPs comes with its intrinsic risks. In order to mitigate these risks as much as possible, backup plans for both observing modes have been integrated into the METIS final design. The backup plan for the vortex observing modes consists of a classical Lyot coronagraph (CLC), while the backup plan for the APP observing mode consists of a binary amplitude apodizer referred to as shaped pupil plate (SPP).\cite{Carlotti11} The design of the Lyot occulting masks (LOM) was optimized together with their dedicated Lyot stop in a trade-off between IWA, throughput, and coronagraphic performance, resulting in an on-sky occulter size of 85~mas for L- and M-band operations, and of 250~mas for N-band operations. The associated Lyot stops, which had to be significantly undersized in their outer edge to mitigate starlight residuals, have a throughput of 63\% compared to the input ELT pupil, leading to an overall throughput around 44\% when taking into account the focal-plane mask throughput and the loss in encircled energy. The design of the SPP was optimized with similar constraints as the vAPP, in a trade-off between IWA, OWA, throughput, and depth of the dark zones. The final design of the SPP, together with the associated PSF, is illustrated in Figure~\ref{fig:spp}. In this design, two half spiders have been made thicker in the pupil stop to enable asymetric pupil focal-plane wavefront sensing\cite{Martinache13} (see Section~\ref{sub:ncpa}). The overall SPP throughput, taking into account the transmission of an AR-coated ZnSe pupil-plane substrate, is about 22\% in terms of encircled energy, compared to the input ELT pupil.

The coronagraphic elements needed to implement the two backup modes will be installed inside the instrument, but will only be tested during assembly, integration and test at system level (system AIT) and deployed on sky in case one of the two HCI baseline modes (vortex and APP) does not deliver according to expectation, due to the intrinsic performance of the as-built coronagraphic components. If the baseline modes are shown to perform as planned during system AIT and commissioning, the backup modes will not be offered to the community nor supported in terms of pipeline processing.

\section{MANUFACTURE, ASSEMBLY, INTEGRATION, AND TEST}

With its system-level final design review (FDR) passed in late 2022, METIS has entered its manufacture, assembly, integration and test (MAIT) phase in earnest. This phase starts at component and subsystem level, followed by the system-level AIT that will consist in integrating and testing all subsystems inside the METIS cryostat at the integration hall in Leiden. Here, we describe the progress with the MAIT of HCI-specific components, which will later be integrated into the appropriate subsystems (CFO, IMG, and LMS).

    \subsection{Vortex phase masks}

\begin{figure}[t]
\centering 
\includegraphics[width=8.5cm]{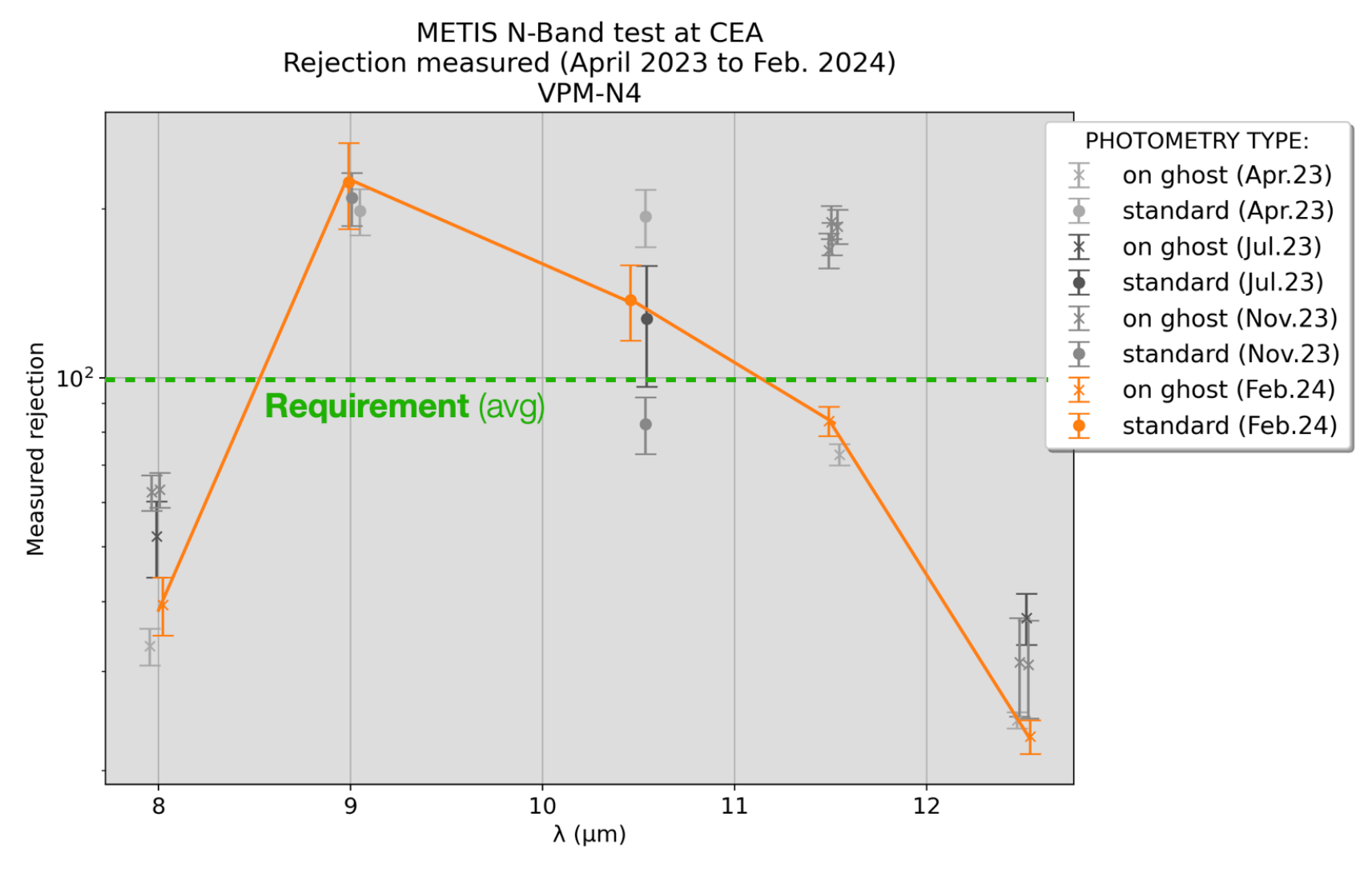}
\includegraphics[width=8.5cm]{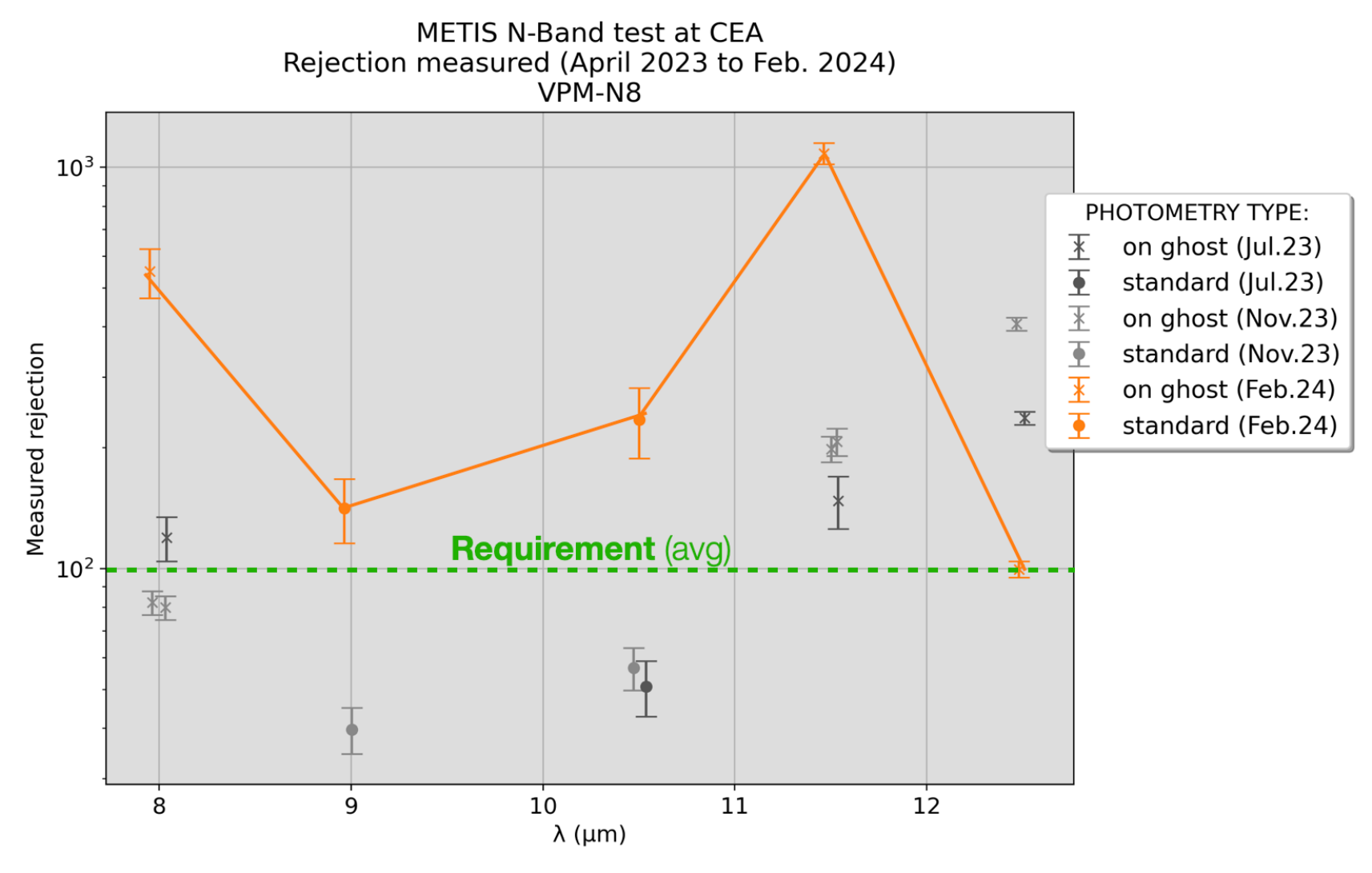}
\caption{Measured performance of two of the three N-band VPMs manufactured for METIS (\textit{left}: VPM-N4, \textit{right}: VPM-N8). Not displayed here is VPM-N9, which shows very similar performance to VPM-N8. The final, consolidated measurements are displayed in orange, with ealier results in shades of gray. In the case of VPM-N8, the gray-colored measurements were obtained before the re-etching of the VPM, showing poor early performance. The requirement in terms of average rejection rate across the N band is displayed with a green dashed line for reference.}
\label{fig:vpm-n}
\end{figure}

At the preliminary design review (PDR), the vortex phase masks (VPM) were identified as long-lead items. They were therefore the subject of an early FDR procedure with a series of other long lead items, which took place in mid-2021. A procurement contract was then signed with Uppsala University in Spring 2022, and the manufacturing started in earnest in Fall 2022. The manufacturing started with the N-band VPMs, because their testing requires more time due to the need for a specific cryogenic facility at CEA Saclay.\cite{Ronayette20} A first N-band VPM, referred to as VPM-N4, was manufactured in February 2023, and --after having to handle some issues with diamond substrate quality-- two additional ones, referred to as VPM-N8 and N9, were manufactured in June 2023. All three VPMs, which have a diameter of 15~mm and include an anti-reflective (AR) grating on their back side, were tested at CEA Saclay in a series of three cryogenic runs, with constant improvements in the coronagraphic test setup. It turned out that VPM-N8 and N9 did not meet the performance requirement at first, while VPM-N4 proved to be compliant with the 100:1 average rejection rate across the N band (see Figure~\ref{fig:vpm-n}). VPM-N8 and N9 were thus sent back to Uppsala for re-etching, based on an updated estimation of the exact grating parameters (especially grating depth, which is impossible to measure in a non-destructive way) informed by the measured coronagraphic performance for a series of five wavelengths across N band. After re-etching in January 2024, which made the grating about 800~nm shallower, both VPMs were re-tested at CEA in February 2024, demonstrating exquisite performance (well beyond the requirement, see Figure~\ref{fig:vpm-n}). All the details of the manufacturing, testing, and re-etching procedures can be found elsewhere.\cite{Forsberg24,Ronayette20} The three N-band VPMs have now been delivered to ULi\`ege, where they will be soon installed and centered into their custom-designed cartridge, before being integrated into the CFO-FP2 focal-plane wheel.

\begin{figure}[t]
\centering 
\includegraphics[width=10cm]{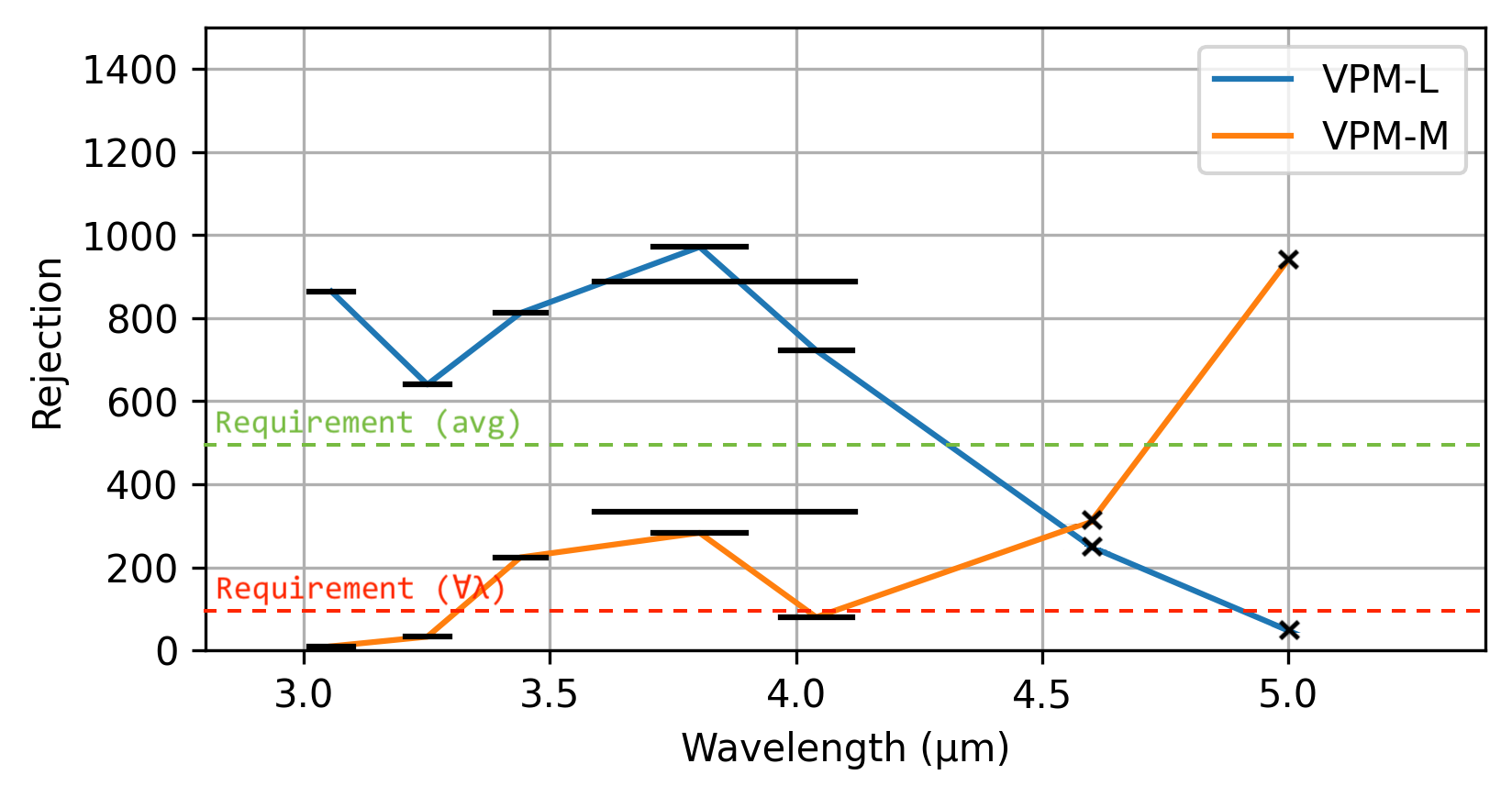}
\caption{Measured performance of VPM-L (blue) and VPM-M (orange), using the VODCA coronagraphic bench of ULi\`ege. The black lines highlight the width of the filters used for the measurements, while the crosses represent monochromatic (QCL laser) measurements. Both have vertical error bars of about 10\% (not shown here). The requirement on the average rejection rate on the operating waveband (500:1) is highlighted in green. The VPMs should also provide a rejection rate better than 100:1 at any given wavelength (red dashed line). These requirements apply to a wavelength range of 2.9--4.1~$\mu$m for VPM-L, and of 3.9--5.3~$\mu$m for VPM-M.\cite{Delacroix24}}
\label{fig:vpm-lm}
\end{figure}

After the N-band VPMs, the L- and M-band VPMs were manufactured in parallel, using 20-mm diameter diamond substrates. The initial etching of the two concentric gratings and of the square AR gratings on their back side was performed in December 2023 at Uppsala, followed by coronagraphic testing on the ULi\`ege VODCA bench\cite{Jolivet19} in February 2024. While VPM-L showed excellent performance, well above the requirement of 500:1 over the 2.9--4.1~$\mu$m range (see Figure~\ref{fig:vpm-lm}), VPM-M showed poor performance across its target operating range (3.9--5.3~$\mu$m), which was expected as the etching process lasted too long, leading to a deeper grating than originally targeted. VPM-M was therefore sent to Uppsala for re-etching, based on a more precise estimation of the grating depth retrieved from the measured rejection rate. VPM-M was then tested again at ULi\`ege in May 2024, showing an average performance compliant with the average rejection rate requirement from about 4.3 to 5.3~$\mu$m (see Figure~\ref{fig:vpm-lm}). While it was originally designed to cover wavelengths down to 3.9~$\mu$m in order to ensure a good overlap with VPM-L, the re-etched version of VPM-M is deemed to be sufficient good for integration into METIS, since the 3.9--4.2~$\mu$m region is very well covered by VPM-L, and since the 4.2--4.4~$\mu$m region is not prone to high-quality ground-based observations due to the very low transparency of the Earth atmosphere at these wavelengths. More details about the MAIT of the METIS vortex phase masks can be found in a dedicated SPIE paper.\cite{Delacroix24}

    \subsection{Apodizing phase plates}

\begin{figure}[t]
\centering 
\includegraphics[width=11.5cm]{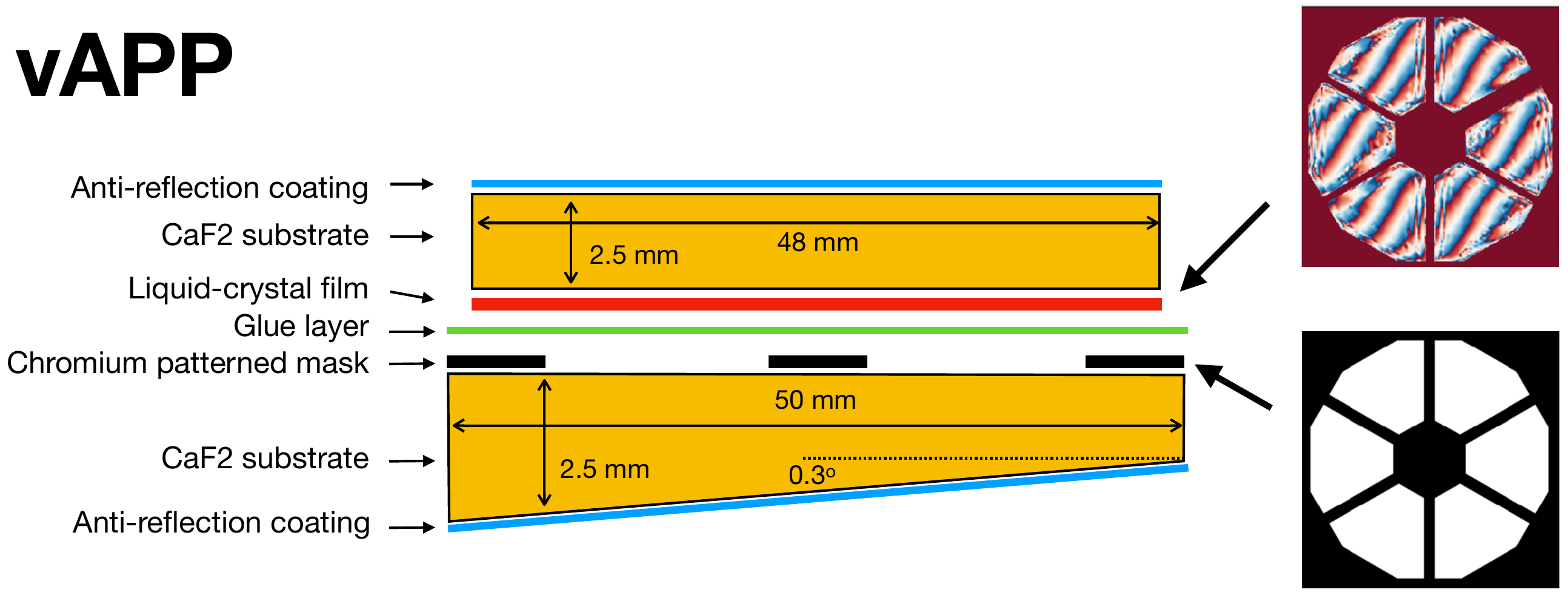} \hspace*{2mm}
\includegraphics[width=5.1cm]{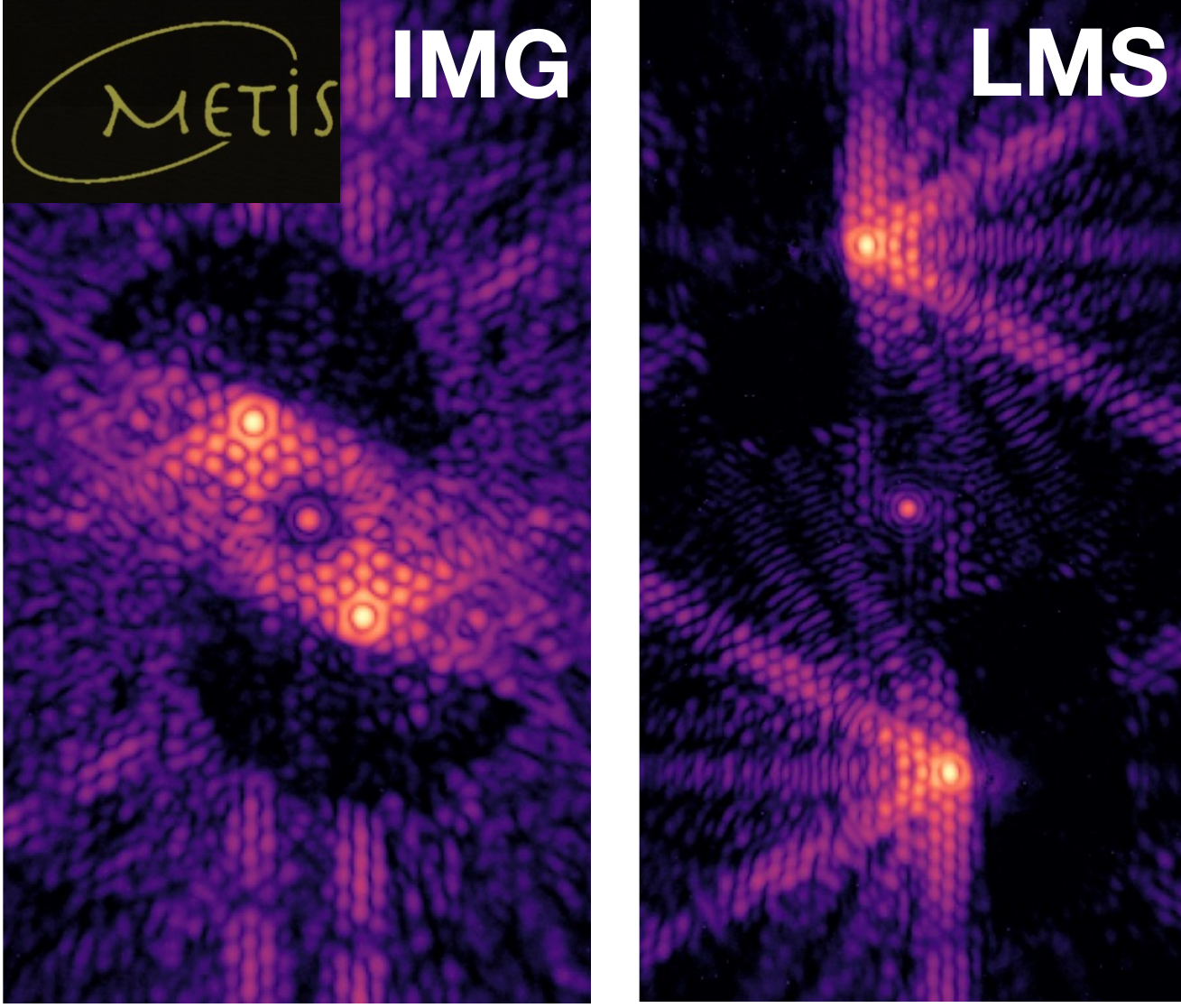}
\caption{\textit{Left.} Practical implementation of the vAPP, illustrated here for APP-IMG. \textit{Right.} In-lab PSFs measured at visible wavelengths for the two vAPP prototypes, respectively for APP-IMG (left) and APP-LMS (right).}
\label{fig:vapp}
\end{figure}

The vAPPs are produced with a liquid-crystal technology, where a direct-write system is used to print the desired fast-axis orientation pattern in a liquid-crystal photo-alignment layer deposited on a substrate. Multiple layers of self-aligning birefringent liquid-crystals are deposited on top of each other with varying thickness and twist, carefully designed to generate the required half-wave retardance.\cite{Doelman21} This liquid crystal layer is then sandwich together with an opaque metal mask between two AR-coated CaF$_2$ substrates, as illustrated in Figure~\ref{fig:vapp}. A wedge is introduced in one of the two substrates to offset any ghost arising from internal reflections inside the substrates. The two substrates are glued together with a very thin, IR-transparent glue layer.

The manufacturing of the vAPPs is currently on-going at ColorLink Japan Ltd. In order to prepare the manufacturing of the final components, several intermediate steps have been taken, including the printing of a visible-wavelength version of the METIS vAPP phase patterns (one for the APP-IMG, one for the APP-LMS), and infrared transmission measurements of the glue layer on a representative CaF$_2$ substrate. The monochromatic PSFs created at visible wavelengths by the two prototypes are illustrated in Figure~\ref{fig:vapp}. These test components demonstrate that the patterning quality is high and that no defects are present in the active areas. Their associated PSFs show a high-quality dark zone except for the presence of an optical ghost not related to the vAPP, and a leakage term of the order of 1\%, demonstrating good control of the liquid-crystal recipes. While the final vAPPs cannot be fully tested in a cryogenic environment ahead of installation into their respective subsystem, we are planning to test the high-contrast performance of APP-LMS on the VODCA bench at ULi\`ege in Fall 2024. This will not be possible for APP-IMG, which has a too large pupil (about 45~mm in diameter compared to 20~mm for APP-LMS) to fit on the VODCA bench.

    \subsection{Other HCI components}

An additional HCI component required a dedicated development: the ring apodizer (RAP). Following previous experiments with similar apodizers\cite{Zhang18,LlopSayson20}, we designed the grayscale region using a microdot pattern. The details related to the design of the microdot pattern are presented in a separate paper.\cite{Konig24} The microdot pattern and the other opaque parts of the RAP (see Figure~\ref{fig:lyot}) are manufactured by Opto-Line Inc through chromium deposition onto an SiO-protected, AR-coated ZnSe substrate, procured from G\&H Artemis Optical. This technology was previously used to manufacture custom Lyot stops for the vortex coronagraph of the NEAR experiment on VLT/VISIR.\cite{Kasper19} Before printing the final component, we printed in Summer 2023 a series of microdot patterns of various sizes, thicknesses, and filling factors, in order to select the microdot parameters that provide the most appropriate effective throughput. These patterns were characterized on the VODCA bench, and tested for robustness to cryogenic conditions. A detailed report on these tests is given elsewhere.\cite{Konig24} Finally, the RAP was printed in June 2024 by Opto-Line Inc, and will be further tested in Li\`ege in Summer 2024.

The coronagraphic elements needed to implement the two backup modes (i.e., the LOMs and SPP) follow the same manufacturing process as the RAP, based on chromium deposition on a SiO-protected, AR-coated ZnSe substrate. The Lyot stops, on the other hand, will be manufactured by drilling holes into a well-chosen substrate, which will then be coated with a black painting such as Acktar Black$^{\rm TM}$.

\section{HCI OPERATION STRATEGY AND SYSTEM-LEVEL TESTING}

In addition to the need for dedicated components, reaching optimal HCI performance also requires specific operation strategies, especially in terms of wavefront control. Here, we summarize our approach to mitigate the effect of non-common path aberrations (NCPA), and briefly describe our plans for testing the METIS HCI performance in representative conditions ahead of commissioning.

    \subsection{NCPA \& focal-plane wavefront sensing} \label{sub:ncpa}

Beyond the design of the coronagraph and the residual wavefront errors from the AO system, one of the main drivers in the performance of any HCI instrument comes from NCPAs.\cite{Guyon18} The requirements on an appropriate mitigation strategy strongly depend on the level and evolution time scale of NCPAs in the instrument. The first order of business is therefore to identify the main sources of NCPAs. The contribution from optical surfaces in the non-common path between the SCAO pyramid wavefront sensor and the first pupil plane inside the imager (where Lyot stops and vAPPs are located) is expected to amount to about 120~nm rms, and to be mostly static since it is fully contained in a gravity-invariant temperature-control vacuum cryostat. The main sources of variable NCPAs are associated with chromatic effects between the near infrared (K band) where SCAO operates, and the mid-infared (L to N bands) where the scientific cameras operate. These chromatic NCPAs have two main origins:
\begin{itemize}
    \item \textit{Chromatic beam wander (CBW)} is due to variations in the optical propagation of slightly offset beams through the METIS optical train, due to differential atmospheric refraction between the operating wavelengths of SCAO and scientific cameras. This contribution has been fully modeled using Fourier optics propagation through the instrument, and will be the subject of a dedicated publication (Bon\'e et al., in prep). This effect creates NCPA variations on a typical timescale of a few minutes, resulting in about 35~nm rms of differential wavefront error variation over an hour.
    \item \textit{Water vapor seeing (WVS)} is related to the strong chromaticity of water vapor at mid-infrared wavelengths, which adds a wavelength-dependent contribution to dry-air seeing. This contribution is described in detail in a previous publication\cite{Absil22}. In terms of wavefront error, it typically ranges from about 30~nm rms at L band to 350~nm rms at N band, with a temporal behavior driven by the wind speed under a frozen flow hypothesis for Kolmogorov-like turbulence.
\end{itemize}

If not corrected, these NCPA variations appearing at typical timescales ranging from a fraction of a second to tens of minutes would severely degrade the HCI performance of METIS, as demonstrated in previous works.\cite{Carlomagno20,Absil22,Delacroix22} This calls for a tailored mitigation strategy. In order to reduce as much as possible the impact of this mitigation strategy on the METIS hardware, we elected to make use of focal-plane wavefront sensing (FP-WFS) based on the data stream from the IMG-LM or IMG-N camera. Non-common path pointing errors at the level of the VPMs will be taken care of with the QACITS algorithm.\cite{Huby15} Higher-order modes will be sensed thanks to a custom implementation of the asymmetric pupil wavefront sensor (AP-WFS)\cite{Martinache13}, which enforces some level of asymmetry in the input pupil to lift the ambiguity on even Zernike modes in the reconstruction of the input wavefront from the measured focal-plane images. While this concept works well for a pupil-plane coronagraph like the vAPP, where one only needs to create a custom asymmetry level in the pupil mask,\cite{Bos19} the situation becomes more complicated for the vortex coronagraph modes, as their performance would be degraded by introducing an asymmetry in the input pupil. Creating such an asymmetry in the input pupil plane would actually not be practical in the case of METIS anyway, as the only available pupil plane upstream of the VPMs (CFO-PP1) is not stabilized in clocking and translation. We will therefore implement the pupil asymmetry downstream of the VPMs, at the level of the Lyot stops, resulting in a new concept referred to as Asymmetric Lyot waveFront sensor (ALF). Because of the nonlinear nature of this FP-WFS concept, the implementation of the wavefront reconstruction will be based on supervised deep learning. This concept is presented in detail in a dedicated paper.\cite{Orban24}

Our two FP-WFS algorithms (QACITS and AP-WFS/ALF) will only be implemented with the IMG cameras. When performing high-contrast observations with the LMS, the structure of the raw data and the associated integration times make it much more challenging to reconstruct the input wavefront in real time. This is one of the reasons why 10\% of the light will be sent to the IMG-LM during LMS observations, allowing to use QACITS for precise centering on the VPM (the same VPM is simultaneously feeding the IMG and LMS). Whether higher-order modes will also be controlled through the AP-WFS/ALF concept during LMS observations will be decided later as a trade-off between the expected performance gain and the operational challenges. In particular, we anticipate that the high-spectral resolution observations obtained with the LMS will not be limited by the same instrumental effects as standard HCI modes, and that the effect of NCPAs might be easier to mitigate in post-processing using techniques based on spectral cross-correlation, for instance\cite{Snellen15}.

    \subsection{System-level testing}

While most of the various HCI components have or will be tested on dedicated test facilities ahead of installation into their respective METIS subsystem, the global performance of the HCI modes can only be assessed once the instrument is fully assembled, including the science cameras (IMG, LMS) that contain among others the vAPPs, Lyot stops, and detectors. These HCI performance tests also require SCAO to work in closed-loop in order to provide a high wavefront quality, which cannot be delivered by the warm calibration unit. This in turn requires a telescope simulator, which needs to include a sufficiently representative deformable mirror, as well as the capabilities to mimic the beam that will be delivered to METIS by the ELT. The design of the METIS telescope simulator, which will be used both for subsystem-level testing of the SCAO module, and for HCI performance testing at system level, is described in a separate paper.\cite{Mortimer24} In addition to testing the system-level performance of the HCI modes, this setup will also be used to validate the two FP-WFS algorithms (QACITS and AP-WFS/ALF), which will be implemented inside the SCAO soft real-time computer.

\section{END-TO-END PERFORMANCE SIMULATIONS} \label{sec:e2e}

Now that the design of the METIS HCI modes has been finalized, and that the performance of the individual HCI components is better understood, we can update the end-to-end performance simulations that were presented just after the preliminary design review.\cite{Carlomagno20}

    \subsection{Sensitivity limits} \label{sub:limits}

Our simulation strategy to estimate the sensitivity limits of METIS was presented in detail in previous publications\cite{Carlomagno20,Delacroix22}, and will not be repeated here. Instead, we will just give a summary of the main steps in our end-to-end simulations. These simulations start by producing a representative 1h-long sequence of SCAO residual phase screens, using a METIS-specific version of the COMPASS simulation software.\cite{Feldt23} This sequence is typically sub-sampled at a frame rate of 300~ms (commensurate with the typical frame rate of IMG-LM), resulting in a total of 12,000 phase screens. These phase screens are then propagated through our instrument simulator using the HEEPS package\cite{Delacroix22}, based on the PROPER Fresnel propagation software.\cite{Krist07} During this propagration process, other instrumental errors such as NCPA (including the variable contributions of CBW and WVS), amplitude errors due to the Talbot effect\cite{Bone20}, pupil drifts upstream of the pupil stabilization mirror, imperfections in the VPM performance, and imperfections in the ELT-M1 pupil (non-uniform reflectivity, or even misaligned segments), are taken into account. This leads to a sequence of 12,000 instantaneous, monochromatic PSFs. This process can be repeated at a series of wavelengths to simulate the effect of the finite spectral filter width, and to include the effect of atmospheric refraction on (differential) pointing errors, in order to build a sequence of polychromatic PSFs. These instantaneous PSFs are then scaled to the appropriate stellar photo-electron rate at the considered detector, and the thermal background is added. We finally add photon noise to these various sources to produce a mock data set, which is then associated with a parallactic angle vector representative of field rotation in a 1h-long observation in pupil tracking mode for a given target. We also produce an off-axis PSF for flux calibration purposes, and present this mock data set to a classical angular differential imaging (cADI) algorithm implemented in the vortex image processing (VIP) package.\cite{Gomez17,Christiaens23} The final product of the VIP-based processing is a $5\sigma$ contrast curve, which represents the sensitivity limit of the observation in terms of planet/star flux ratio.

\begin{figure}[t]
\centering 
\includegraphics[width=\textwidth]{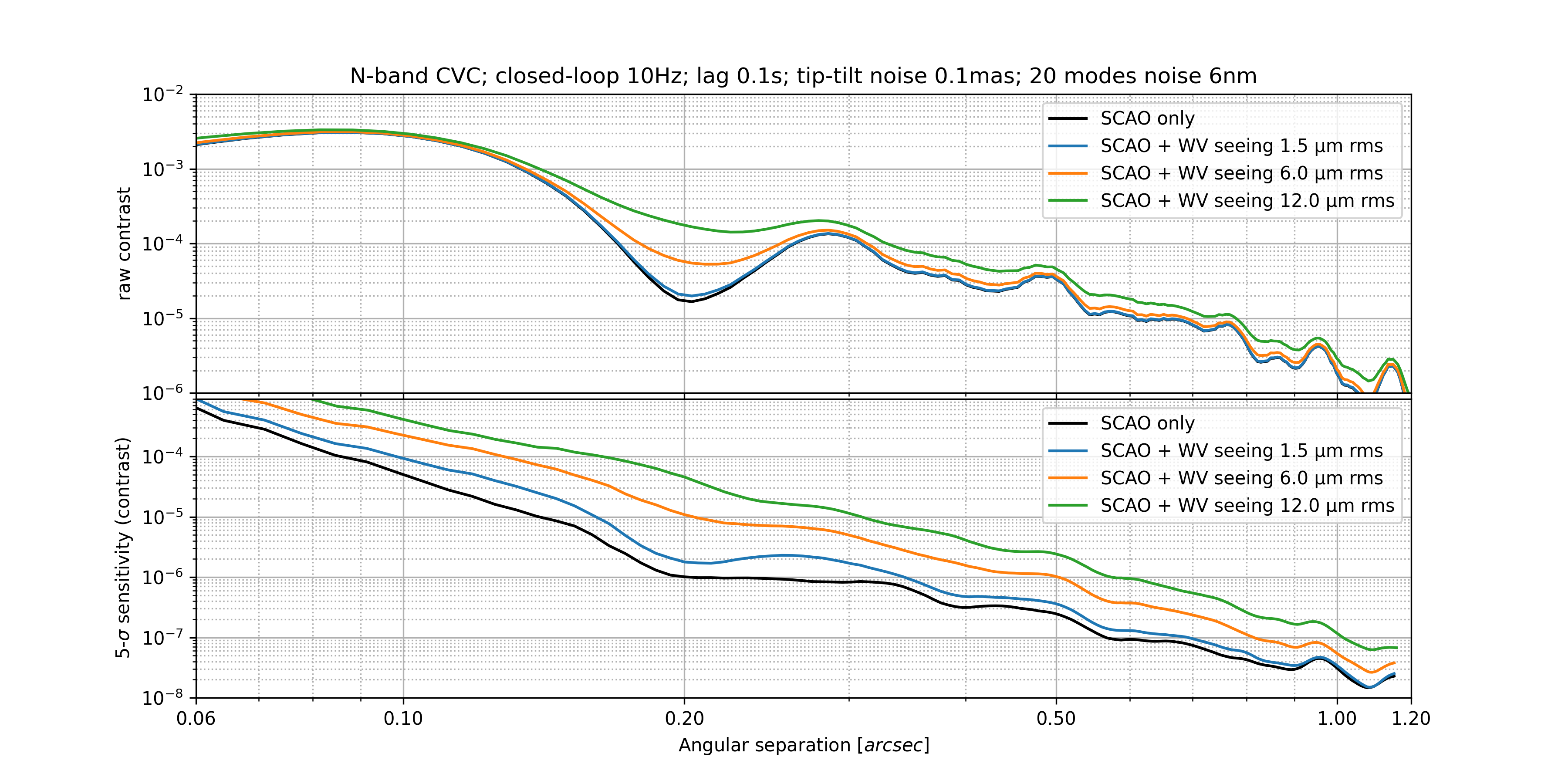}
\caption{Raw contrast (\textit{top}) and post-ADI contrast curves (\textit{bottom}) for the classical vortex coronagraphy mode at N band, under various levels of water vapor seeing (colored lines), compared to SCAO-only simulations (black line). Wavefront correction based on FP-WFS (QACITS and ALF algorithms) is performed on the first 20 Zernike modes at a loop repetition rate of 10~Hz, assuming a realistic noise level in both sensors.}
\label{fig:wvs}
\end{figure}

An important feature in our simulations, which was already introduced in a previous publication\cite{Delacroix22}, is our implementation of the effect of closed-loop wavefront control based on our FP-WFS algorithms. For both CBW and WVS, we generate representative time sequences of NCPA phase screens, which we feed to a time-domain closed-loop simulation that takes into account the estimated noise in the FP-WFS as a function of the input star magnitude. These estimations of the actual amount and variability of residual NCPA at the level of the coronagraph are key to obtain a realistic prediction of the post-ADI sensitivity limits, as NCPAs are expected to be one of the main drivers in the HCI performance budget.\cite{Carlomagno20} In this context, it is particularly interesting to investigate the influence of the strength of WVS on the HCI performance at N band, where WVS largely dominates the performance budget. Based on archival data collected with the GRAVITY fringe tracker at the Very Large Telescope Interferometer (VLTI), we have been able to obtain a first rough estimate of the variability in the strength of WVS, using the strategy presented in a previous work.\cite{Absil22} Based on this preliminary analysis, the median conditions are expected to correspond to about 6~$\mu$m of rms differential precipitable water vapor (PWV) on a 30-m baseline, which corresponds to an rms variation in differential WV column density of about $2\times 10^{19}$~cm$^{-2}$. These median conditions lead to an additional wavefront error of about 30~nm rms at L band and about 350~nm rms at N band, as already mentioned in Sect.~\ref{sub:ncpa}. The best WVS conditions (percentile 10) may be as good as 1.5~$\mu$m rms differential PWV (i.e., four times lower variability than the median conditions), while the worst conditions (percentile 90) are estimated to be around 12~$\mu$m rms differential PWV (i.e., twice as much as the median conditions). These preliminary values still need to be confirmed with more data, and will be fully reported in a future publication. Based on these preliminary estimates, one can update the end-to-end simulations at N band, taking into account the closed-loop correction provided by QACITS and ALF. The results of these simulations are displayed in Figure~\ref{fig:wvs}, where we assume a 10-Hz loop repetition rate, which is expected to be achievable on bright targets (N-band magnitude lower than $N\simeq 2$). These simulations show how critical it will be to identify the episodes of lowest WVS to reach the best possible HCI performance at N band. During these episodes, the HCI performance could be close to being SCAO-limited -- not considering other instrumental effects, such as CBW or amplitude aberrations due to the Talbot effect, which may start to dominate the error budget. It is expected that these episodes of very low WVS are correlated with the PWV content in the atmosphere, as well as to dry air seeing to some level, although the statistical behavior and correlation of WVS with respect to other atmospheric effects still need to be further studied.

Our latest estimation of the L- and M-band HCI performance in terms of post-ADI contrast curves does not significantly differ from our 2022 SPIE publication\cite{Delacroix22}, as WVS seeing is not the main contribution in the error budget at those wavelengths, and therefore need not be repeated here.

    \subsection{Prospects for high-contrast exoplanet science}

\begin{figure}[t]
\centering 
\includegraphics[width=8cm]{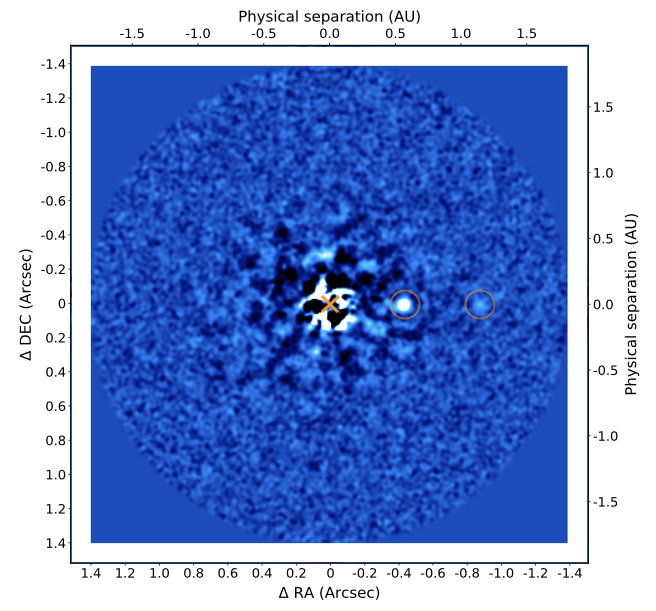} \hspace*{2mm}
\includegraphics[width=8cm]{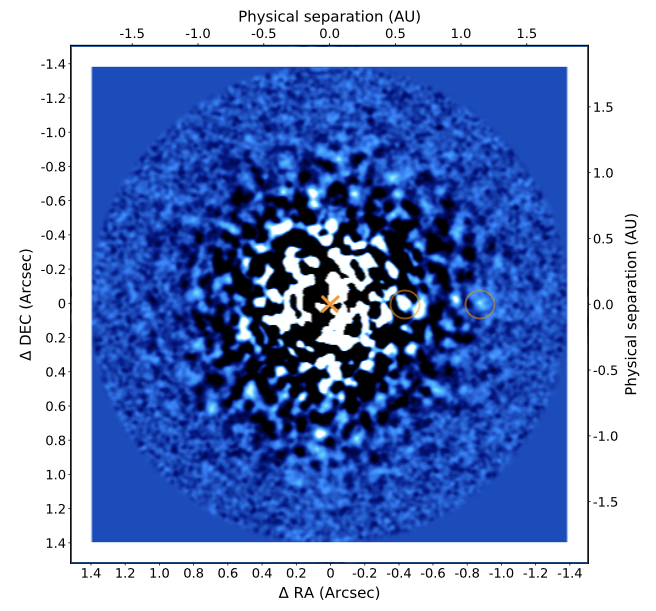}
\caption{End-to-end simulations of a 5h ADI observing sequence on $\alpha$~Cen~A at N band with the classical vortex coronagraph. Two Earth-like planets (i.e., one Earth radius with an Earth-like atmosphere) were injected into these simulations (orange circles), one at 1.1~au where it would receive the same irradiation as the Earth from the Sun, and another (warmer and hence brighter) one at half that separation. \textit{Left.} Post-ADI image in the absence of WVS, which is representative of the best WVS conditions that should be reachable at the ELT site. \textit{Right.} Post-ADI image assuming the worst WVS conditions, as described in Sect.~\ref{sub:limits}.}
\label{fig:alfcen}
\end{figure}

With its high-contrast and high-spectral resolution capabilities, METIS is expected to enable new breakthroughs in the field of exoplanet science, among others. It is beyond the scope of this paper to provide a comprehensive overview of all the HCI-related science cases of METIS. Yet, based on the new N-band simulations presented above, it is useful to reconsider the capability of METIS to detect potential Earth-like planets around our nearest neighbors, and around $\alpha$~Cen~A in particular, which is the most promising target for N-band imaging of an Earth twin.\cite{Bowens21} The final ADI images obtained for a 5-h ADI observing sequence with two extreme levels of WVS are displayed in Figure~\ref{fig:alfcen}, where we injected two Earth twins respectively in the middle of the habitable zone (0.82'' separation at quadrature), and twice closer. In the absence of WVS, the two planets are recovered with an SNR of 6 and 11, respectively, while both of them are below the detection threshold ($\text{SNR}=5$) in the case of strong WVS. These simulations highlight two important points: (a) in the absence of WVS, the habitable zone around $\alpha$~Cen~A lies in the background-limited and not contrast-limited regime, and (b) METIS has the potential of imaging an Earth twin within a 5h integration time, under excellent WVS conditions. METIS also has the potential of revealing warmer rocky planets at N band around this most favorable target star, even though in general warmer planets may benefit from observations at shorter wavelengths (L and M bands), which also provide a significant boost in angular resolution.

\section{CONCLUSION \& PERSPECTIVES}

Over the last couple of years, the METIS project has moved from the final design phase to the manufacture, assembly, integration, and test phase. The various components dedicated to the implementation of the HCI modes are now either fully manufactured, or making good progress towards being manufactured and qualified for integration into METIS. In the coming months, the HCI activities will therefore focus more on the finalization of HCI-specific software for focal-plane wavefront sensing, and of its integration into the adaptive optics control architecture. Finally, once METIS is fully assembled in the integration hall at Leiden (2027--2028), the HCI activities will focus on the validation of the HCI modes at system level, including end-to-end performance tests under closed-loop adaptive optics conditions, and including system-level testing of the FP-WFS algorithms.

\acknowledgments 
 
This project has received funding from the European Research Council (ERC) under the European Union's Horizon 2020 research and innovation programme (grant agreement No 819155), and from the Wallonia-Brussels Federation (grant for Concerted Research Actions).

\bibliography{report} 
\bibliographystyle{spiebib} 

\end{document}